\documentstyle[sprocl]{article}
\begin{document}
\title{NEARLY DEGENERATE NEUTRINO MASSES AND NEARLY 
DECOUPLED NEUTRINO OSCILLATIONS
\footnote{Talk given by one of us (H.F.) in the 17th 
International Workshop on Weak Interactions and
Neutrinos, Cape Town, South Africa, January 1999}}
\author{{\bf Harald Fritzsch} ~ and ~ {\bf Zhi-zhong Xing}}
\address{
Sektion Physik, Universit$\ddot{a}$t M$\ddot{u}$nchen,
80333 M$\ddot{u}$nchen, Germany
}
\maketitle
\abstracts{
We introduce a simple flavor symmetry breaking scheme, in which
charged lepton masses have a strong hierarchy and neutrino masses 
are almost degenerate. It is possible to obtain a natural
suppression of lepton flavor mixing between the 1st
and 3rd families as well as 
the approximate decoupling of atmospheric and solar neutrino
oscillations with nearly bi-maximal mixing factors.
The similarity and difference between lepton and quark flavor
mixing schemes are briefly discussed.
}

In the standard model neutrinos are assumed to be the 
massless Weyl particles. But most extentions of
the standard model (such as the grand unified theories of quarks
and leptons) allow the existence of massive neutrinos, although the
masses of three active neutrinos $\nu_e$, $\nu_\mu$ and $\nu_\tau$
could be much smaller than those of their
charged counterparts. Whether the smallness of the masses of three
neutrinos are attributed to the neutrality of their electric 
charges or to the Majorana feature of their fields, remains an 
open question. 

The recent observation of the
atmospheric and solar neutrino anomalies, particularly
that in the Super-Kamiokande experiment, has provided 
strong evidence that neutrinos are
massive and lepton flavors are mixed \cite{Suzuki99}. 
Analyses of the atmospheric neutrino deficit in the framework
of two-flavor neutrino oscillations yield the following
mass-squared difference and mixing factor:
\begin{equation}
\Delta m^2_{\rm atm} \; \sim \; 10^{-3} ~ {\rm eV^2} \; ,
~~~~~~~~
\sin^2 2\theta_{\rm atm} \; > \; 0.8 \; .  
\end{equation}
In addition, the hypothesis that solar $\nu_e$ neutrinos
change to another active species through
long-wavelength vacuum oscillations with the parameters
\begin{equation}
\Delta m^2_{\rm sun} \; \sim \; 10^{-10} ~ {\rm eV^2} \; ,
~~~~~~~~~~
\sin^2 2\theta_{\rm sun} \; \approx \; 1 \; ,
\end{equation}
can provide a consistent
explanantion of all existing solar neutrino data \cite{Barger99}.
In the framework of three-flavor neutrino
oscillations, the significant hierarchy between $\Delta m^2_{\rm atm}$
and $\Delta m^2_{\rm sun}$ together with the no observation
of $\bar{\nu}_e \rightarrow \bar{\nu}_e$ oscillation in the
CHOOZ experiment implies that the $\nu^{~}_3$-component in $\nu_e$
is rather small (even negligible) and the atmospheric and
solar neutrino oscillations approximately decouple \cite{Giunti98}. 
If this picture is true, then the solar and
atmospheric neutrino deficits should mainly attributed to 
the corresponding $\nu_e \rightarrow \nu_\mu$ and $\nu_\mu \rightarrow
\nu_\tau$ transitions. In this case, 
\begin{eqnarray}
\Delta m^2_{\rm sun} & = & \Delta m^2_{21} \; \equiv \;
\left | m^2_2 - m^2_1 \right | \; , \nonumber \\
\Delta m^2_{\rm atm} & = & \Delta m^2_{32} \; \equiv \;
\left | m^2_3 - m^2_2 \right | \; ,
\end{eqnarray}
and $\Delta m^2_{31} \approx \Delta m^2_{32}$. Nevertheless, the
hierarchy of $\Delta m^2_{21}$ and $\Delta m^2_{32}$ (or 
$\Delta m^2_{31}$) does not give information about the absolute values
or the relative magnitude of three neutrino masses. For example,
either the strongly hierarchical neutrino mass spectrum 
($m_1 \ll m_2 \ll m_3$) or the nearly degenerate one 
($m_1 \approx m_2 \approx m_3$) is allowed to reproduce the 
``observed'' gap between $\Delta m^2_{21}$ and $\Delta m^2_{32}$.

In this talk we pay attention only to the mass degeneracy of 
active neutrinos, which might be good candidates for the hot
dark matter of the universe. We 
introduce a simple flavor symmetry breaking 
scheme for charged lepton and neutrino mass matrices, so as to
generate two nearly bi-maximal flavor mixing angles and to
interpret the approximate decoupling of solar and 
atmospheric neutrino oscillations. Within the scope of this 
discussion, we do not take 
the LSND evidence for neutrino oscillations and
the matter-enhanced mechanism for solar neutrino oscillations
into account.

Let us start with the symmetry limits of the charged lepton 
and neutrino mass matrices. In a
specific basis of flavor space, in which charged leptons have the
exact flavor democracy and neutrino masses are fully degenerate,
the mass matrices can be written as \cite{FX96,FX98}
\begin{equation}
M^{(0)}_l \; =\; \frac{c^{~}_l}{3} \left (\matrix{
1       & 1     & 1 \cr
1       & 1     & 1 \cr
1       & 1     & 1 \cr} \right ) \; , ~~~~~~~~
M^{(0)}_\nu \; =\; c_\nu \left (\matrix{
1       & 0     & 0 \cr
0       & 1     & 0 \cr
0       & 0     & 1 \cr} \right ) \; ,
\end{equation}
where $c^{~}_l =m_\tau$ and $c_\nu =m_0$ measure the 
corresponding mass scales.
If the three neutrinos are of the Majorana type and $CP$ symmetry is
conserved, $M^{(0)}_\nu$ could take a more general form
$M^{(0)}_\nu P_\nu$, where $P_\nu = {\rm Diag} \{ \eta_1,
\eta_2, \eta_3 \}$ with $\eta_i= \pm 1$ 
denoting the $CP$ parities. For simplicity we neglect the effect
of $P_\nu$, which is only relevant to the neutrinoless double
beta decay, in the subsequent discussions.
Clearly $M^{(0)}_\nu$ exhibits an
S(3) symmetry, while $M^{(0)}_l$ an
$S(3)_{\rm L} \times S(3)_{\rm R}$ symmetry.
In these limits $m_e = m_\mu =0$,
$m_1 = m_2 =m_3 =m_0$, and no flavor mixing is present.

A simple diagonal breaking of the flavor democracy
for $M^{(0)}_l$ and the mass degeneracy for $M^{(0)}_\nu$
may lead to instructive results for 
neutrino oscillations \cite{FX96}. 
Let us proceed with two different symmetry-breaking steps.

(i) Small perturbations to the (3,3) elements of $M^{(0)}_l$
and $M^{(0)}_\nu$ are respectively introduced \cite{FH94}:
\begin{equation}
\Delta M^{(1)}_l \; = \; \frac{c^{~}_l}{3} \left ( \matrix{
0       & 0     & 0 \cr
0       & 0     & 0 \cr
0       & 0     & \varepsilon^{~}_l \cr } \right ) \; , 
~~~~~~
\Delta M^{(1)}_\nu \; = \; c_\nu \left ( \matrix{
0       & 0     & 0 \cr
0       & 0     & 0 \cr
0       & 0     & \varepsilon_\nu \cr } \right ) \; .
\end{equation}
In this case the charged lepton mass matrix $M^{(1)}_l =
M^{(0)}_l + \Delta M^{(1)}_l$ remains symmetric under an
$S(2)_{\rm L}\times S(2)_{\rm R}$ transformation, 
and the neutrino mass matrix
$M^{(1)}_\nu = M^{(0)}_\nu + \Delta M^{(0)}_\nu$ has
an $S(2)$ symmetry. 
The muon becomes massive ($m_\mu \approx 2|\varepsilon^{~}_l|
m_\tau /9$), and the mass eigenvalue $m_3$ is no more degenerate
with $m_1$ and $m_2$ (i.e., $|m_3 - m_0| = m_0 |\varepsilon_\nu|$). 
After the diagonalization of 
$M^{(1)}_l$ and $M^{(1)}_\nu$, one finds that the 2nd and 3rd
lepton families have a definite flavor mixing angle
$\theta$. We obtain $\tan\theta = -\sqrt{2} ~$ if the
small correction of $O(m_\mu/m_\tau)$ is neglected.
Then neutrino oscillations at the atmospheric scale may arise
in $\nu_\mu \rightarrow \nu_\tau$ transitions with 
$\Delta m^2_{32} = \Delta m^2_{31}
\approx 2m_0 |\varepsilon_\nu|$. The corresponding
mixing factor $\sin^2 2\theta \approx 8/9$ is in good agreement 
with current data \cite{Suzuki99}.

(ii) Small perturbations, which have the identical magnitude but
the opposite signs \cite{FX96}, 
are introduced to the (2,2) and (1,1) elements of
$M^{(1)}_l$ or $M^{(1)}_\nu$:
\begin{equation}
\Delta M^{(2)}_l = \frac{c^{~}_l}{3} \left ( \matrix{
-\delta_l       & 0     & 0 \cr
0       & \delta_l      & 0 \cr
0       & 0     & 0 \cr } \right ) \; , 
~~~~~
\Delta M^{(2)}_\nu = c_\nu \left ( \matrix{
-\delta_\nu     & 0     & 0 \cr
0       & \delta_\nu    & 0 \cr
0       & 0     & 0 \cr } \right ) \; .
\end{equation}
We obtaine $m_e \approx |\delta_l|^2 m^2_\tau /(27 m_\mu)$ 
and $m_1 = m_0 (1-\delta_\nu)$,
$m_2 = m_0 (1+\delta_\nu)$. The diagonalization of 
$M^{(2)}_l = M^{(1)}_l + \Delta M^{(2)}_l$ and 
$M^{(2)}_\nu = M^{(1)}_\nu + \Delta M^{(2)}_\nu$ 
leads to a full $3\times 3$
flavor mixing matrix, which links neutrino mass eigenstates 
$(\nu_1, \nu_2, \nu_3)$ to neutrino flavor eigenstates
$(\nu_e, \nu_\mu, \nu_\tau)$ in the following manner:
\begin{equation}
V \; =\; \frac{1}{\sqrt{6}} \left ( \matrix{
\sqrt{3}        & -\sqrt{3}     & 0 \cr
1       & 1     & -2 \cr
\sqrt{2}        & \sqrt{2}      & \sqrt{2} \cr}
\right ) 
\; \pm \; \xi^{~}_V ~ \sqrt{\frac{m_e}{m_\mu}} 
\; \mp \; \zeta^{~}_V ~ \frac{m_\mu}{m_\tau} \; ,
\end{equation}
where
\begin{equation}
\xi^{~}_V = \frac{1}{\sqrt{6}} \left ( \matrix{
1       & 1     & -2 \cr
-\sqrt{3}       & \sqrt{3}      & 0 \cr
0       & 0     & 0 \cr} \right ) \; ,
~~~
\zeta^{~}_V = \frac{1}{2\sqrt{3}} \left ( \matrix{
0       & 0     & 0 \cr
\sqrt{2}        & \sqrt{2}      & \sqrt{2} \cr
-1      & -1    & 2 \cr}
\right ) \; .
\end{equation}
Some comments on this result are in order.
\begin{itemize}
\item   This mixing pattern, after neglecting small
corrections from the charged lepton masses, is similar to that
of the pseudoscalar mesons $\pi^0$, $\eta$ and $\eta'$ in 
QCD \cite{FH94,F98}:
\begin{eqnarray}
|\pi^0\rangle & = & \frac{1}{\sqrt{2}} \left (
|\bar{u}u\rangle - |\bar{d}d\rangle \right ) \; ,
\nonumber \\
|\eta\rangle & = & \frac{1}{\sqrt{6}} \left (
|\bar{u}u\rangle + |\bar{d}d\rangle - 2 |\bar{s}s\rangle
\right ) \; ,
\nonumber \\
|\eta'\rangle & = & \frac{1}{\sqrt{3}} \left (
|\bar{u}u\rangle + |\bar{d}d\rangle + |\bar{s}s\rangle 
\right ) \; .
\end{eqnarray}
One is invited to speculate whether such an analogy could 
be taken as a hint towards an underlying symmetry responsible for
lepton mass generation \cite{TH}.

\item   The (1,3) element of $V$ is naturally suppressed in
the symmetry breaking scheme outlined above. A similar 
feature appears in the $3\times 3$ quark flavor mixing
matrix, i.e., $|V_{ub}|$ is the smallest among the
nine quark mixing elements. Indeed the smallness of $V_{e3}$
provides a necessary condition for the decoupling of
solar and atmospheric neutrino oscillations, even though neutrino
masses are nearly degenerate. The effect of small but nonvanishing
$V_{e3}$ can manifest itself in the long-baseline $\nu_\mu
\rightarrow \nu_e$ and $\nu_e \rightarrow \nu_\tau$ transitions,
as we shall see below.

\item   The flavor mixings between the 1st and 2nd lepton families
and between the 2nd and 3rd lepton families are nearly
maximal. This property, together with the natural smallness
of $V_{e3}$, allows a satisfactory interpretation of the 
observed large mixing 
in atmospheric and solar neutrino oscillations. We obtain 
\begin{equation}
\sin^2 2\theta_{\rm sun} \; = \; 1 \; , ~~~~~~~~
\sin^2 2\theta_{\rm atm} \; =\; \frac{8}{9} \left ( 1 \mp
\frac{m_\mu}{m_\tau} \right ) \; 
\end{equation}
to a high degree of accuracy. At present both solutions for 
$\sin^2 2\theta_{\rm atm}$, i.e., $\sin^2 2\theta_{\rm atm}
= 0.84$ or $0.94$, are allowed by data.
\end{itemize}

Let us make a brief but useful comparison
between the lepton and quark flavor mixing schemes. 
For simplicity we make use of the following 
parametrization \cite{FX97}:
\begin{equation}
V \; =\; \left ( \matrix{
s^{~}_x s_y c + c^{~}_x c_y     e^{-{\rm i}\phi}
& s^{~}_x c_y c - c^{~}_x s_y e^{-{\rm i}\phi}
& s^{~}_x s \cr
c^{~}_x s_y c - s^{~}_x c_y e^{-{\rm i}\phi}    
& c^{~}_x c_y c + s^{~}_x s_y e^{-{\rm i}\phi}
& c^{~}_x s \cr
-s_y s  & -c_y s        & c \cr } \right ) \; . 
\end{equation}
For leptons we take the subscripts $x =l$ and $y = \nu$,
while for quarks $x = \rm u$ and $y = \rm d$. Therefore 
the rotation angle $\theta_l$ (or $\theta_\nu$) mainly
describes the mixing between $e$ and $\mu$ leptons (or between
$\nu_e$ and $\nu_\mu$ neutrinos), and the rotation angle $\theta_{\rm u}$
(or $\theta_{\rm d}$) primarily describes the mixing between
$u$ and $c$ quarks (or between $d$ and $s$ quarks). The rotation
angle $\theta$ is a combined effect arising from the mixing between
the 2nd and 3rd families, for either quarks or leptons. The
phase parameter $\phi$ signals $CP$ violation in flavor mixing
(for neutrinos of the Majorana type, two additional $CP$-violating
phases may enter but they are irrelevant for neutrino oscillations).
Comparing Eqs. (7) and (11) we immediately arrive at
(up to a sign ambiguity of $\theta_l$) 
\begin{equation}
\tan\theta_l \; =\; \sqrt{\frac{m_e}{m_\mu}} \; ,
~~~~~~~~ \tan\theta_\nu \; =\; 1 \; . \;\;\;\;\;
\end{equation}
In contrast, a variety of quark mass matrices predict \cite{F78,F99}
\begin{equation}
\tan\theta_u \; =\; \sqrt{\frac{m_u}{m_c}} \; ,
~~~~~~~~ \tan\theta_d \; =\; \sqrt{\frac{m_d}{m_s}} \; .
\end{equation}
As one can see, the large mixing angle $\theta_\nu$ is attributed
to the near degeneracy of neutrino masses in our flavor symmetry
breaking scheme. 

Finally we consider the effect of nonvanishing $\theta_l$ for
$\nu_\mu \rightarrow \nu_e$ and
$\nu_e \rightarrow \nu_\tau$ transition probabilities in 
long-baseline (LB) neutrino experiments, in which the oscillations
associated with the mass-squared difference $\Delta m^2_{21}$
can safely be neglected. We obtain \cite{FX98}
\begin{eqnarray}
P(\nu_\mu \rightarrow \nu_e)^{~}_{\rm LB} & = & 
\frac{16}{9} \frac{m_e}{m_\mu}
\sin^2 \left ( 1.27 ~ \frac{\Delta m^2_{32} L}{|{\bf P}|} \right ) \; ,
\nonumber \\
P (\nu_e \rightarrow \nu_\tau)_{\rm LB} & = &
\frac{~8~}{9} \frac{m_e}{m_\mu} 
\sin^2 \left ( 1.27 ~ \frac{\Delta m^2_{32} L}{|{\bf P}|} \right ) \; .
\end{eqnarray}
The mixing factors in these two processes are $0.8\%$ and $0.4\%$,
respectively. The former might be within the sensitivity region of
MINOS.

More data from the Super-Kamiokande and other neutrino
experiments could finally clarify whether the
solar neutrino deficit is attributed to the long-wavelength
vacuum oscillation. They will provide stringent tests of the
model discussed here.

\end{document}